\documentclass[epsfig,12pt,onecolumn]{article}
\usepackage{amsfonts}
\usepackage{amssymb}
\usepackage{multicol}
\usepackage{graphicx}
\usepackage{float}
\usepackage{caption}
\usepackage{xcolor}
\usepackage[utf8]{inputenc}
\usepackage{amsmath}

\makeatletter \@addtoreset{equation}{section}
\renewcommand{\theequation}{\thesection.\arabic{equation}}
\def \be{\begin{equation}}
\def \ee{\end{equation}}
\def \bea{\begin{eqnarray}}
\def \eea{\end{eqnarray}}

\newcommand{\nc}{\newcommand}
\nc{\al}{\alpha} \nc{\bib}{\bibitem} \nc{\la}{\lambda}
\nc{\C}{\mbox{\hspace{1.24mm}\rule{0.2mm}{2.5mm}\hspace{-2.7mm} C}}
\nc{\R}{\mbox{\hspace{.04mm}\rule{0.2mm}{2.8mm}\hspace{-1.5mm} R}}
\input{tcilatex}
\begin{document}

\title{\textbf{Landscape of Narain CFTs}}
\author{R. Sammani$^{1,2}$, E.H Saidi$^{1,2,3}$, R. Ahl Laamara$^{1,2}$, L.B
Drissi$^{1,2,3}$ \\
{\small 1. LPHE-MS, Science Faculty}, {\small Mohammed V University in
Rabat, Morocco.}\\
{\small 2. Centre of Physics and Mathematics, CPM- Morocco.}\\
{\small 3. Hassan II Academy of Science and Technology, Kingdom of Morocco.}}
\maketitle

\begin{abstract}
In this work, we investigate the AdS$_{3}$ gravitational bulk dual to an
ensemble of Narain CFTs and their generalisations to establish bounds
consistent with the Swampland program. Focusing on the AdS distance and
finiteness conjectures, we show that the central charge of Narain CFTs
forming the ensemble must be\ finite. Combining anomaly and unitary
requirements, we derive an upper bound on the rank of the abelian U(1) gauge
symmetries that can consistently couple to the AdS$_{3}$ gravity. We give
explicit realisations of these constraints by determining the range of the
Chern-Simons level $k^{G}$ corresponding to a
bounded AdS$_{3}$ radius$.$ Accordingly, the Narain landscape is finite with
a number of admissible CFTs constrained as $3/2\lesssim c\lesssim 10^{3}.$ \\
\textbf{Keywords:} Narain CFTs, Swampland program, Holography, AdS$_{3}$/CFT$%
_{2}$, Anomaly cancellation.
\end{abstract}

\section{Introduction}

The AdS/CFT correspondence has long provided a powerful framework to study
quantum gravity by connecting gravitational theories in Anti de Sitter (AdS)
space to conformal field theories (CFT) on their boundary \cite{AdSCFT}.
More recently, the development of the Swampland program has deepened our
understanding of the constraints that consistent effective field theory must
satisfy to be UV completed into quantum gravity theories \cite{SP1}-\cite%
{SP4}. By merging insights from both frameworks, we examine and constrain
the structure of the Landscape of admissible CFT$_{2}$ dual to consistent
gravitational theories in AdS$_{3}$ backgrounds.

We are interested in an ensemble of Narain CFTs \cite{MW, Meer1, Meer2} as
the boundary dual of AdS$_{3}$ gravity coupled to abelian U(1) gauge fields.
It has been demonstrated that such ensembles admit a gravitational
description in terms of an exotic U(1)$^{2p}$ gravity, which itself can be
viewed as an AdS$_{3}$ gravity with isometry group $SO(2,2)=SL(2,R)_{L}%
\times SL(2,R)_{R}$ coupled to $U(1)_{L}^{p}\times U(1)_{R}^{p}$ gauge
fields, provided that suitable diagonal boundary conditions are imposed \cite%
{PT, Fluc}. Our purpose in this paper is to derive Swampland consistency
conditions on the gravitational bulk to delineate the Landscape of Narain
CFTs and their generalisations.

Considering an ensemble of Narain CFTs, upon averaging over the Narain
moduli space, global symmetries emerge which are forbidden by the
Swampland's no global symmetry conjecture. To restore consistency, one must
introduce fluctuations given by a shift between the Narain theta function
and the Eisenstein series describing the averaged partition function \cite%
{M3}. These fluctuations have been shown to correspond to a sublattice of
superextremal particles emitted by the discharging BTZ black hole \cite{Fluc}%
. This aligns with the weak gravity conjecture and its sublattice refinement
\cite{wgc3}-\cite{MI}. In this work, we turn our attention to two additional
Swampland criteria: the AdS distance conjecture \cite{DC} and the finiteness
of the Landscape conjecture \cite{Finit1}-\cite{Finit4}.

For large AdS radius, the AdS distance condition predicts the emergence of
an infinite tower of states \cite{DC}. The finiteness conjecture on the
other hand, asserts that only a finite number of consistent AdS and their
corresponding CFTs can exist within the Landscape \cite{Finit1}-\cite{rev}.
In this paper, we aim to investigate the interplay between these two
conjectures in the context of AdS$_{3}$ theories with an ensemble of Narain
CFTs at their boundary and demonstrate how bulk gravitational consistency
conditions can impose constraints on the CFT ensemble. Additionally, we
consider recent generalisations of Narain theories to broaden the scope of
our results and reinforce our interpretations and conclusions.

We first study the emerging anomalies in the bulk gravity that we split into
two sectors: $\left( \mathbf{i}\right) $ a pure gravitational sector
formulated as a Chern-Simons (CS) action with $SL(2,R)_{L}\times SL(2,R)_{R}$
symmetry; and $\left( \mathbf{ii}\right) $ a gauge sector with $%
U(1)_{L}^{p}\times U(1)_{R}^{p}$ abelian fields. We show that the pure
gravitational sector remains healthy of anomalies under Grumiller-Riegler's
boundary conditions \cite{GR} provided that the central charges are equal.
In contrast, the gauge sector, for both Narain and extended Narain CFTs,
exhibits gauge anomalies and in the case of generalised Narain CFTs,
gravitational anomalies as well. After introducing two cancellations
mechanisms, one to address the gauge anomalies and another for the
gravitational ones, we derive an upper bound on the rank of the abelian U(1)
gauge symmetries that can consistently couple to the AdS$_{3}$ gravity. We
provide an explicit realisation of the finiteness and unitarity bounds by
leveraging dimensional analysis, geometrical conditions and insights from
numerical bootstrap program \cite{SB1}-\cite{Bos1}. This permits to
constraint the CS level as $1/4\lesssim k^{G}\lesssim 166$ and the AdS$_{3}$
radius like $G_{N}\lesssim l_{AdS_{3}}\lesssim 666G_{N}$ which enables the
computation of the boundaries of the Narain Landscape leading us to conclude
that the number of CFTs forming the Narain landscape is finite.

The organisation of this paper is as follows:\ In section 2, we overview the
construction of Narain CFTs, their extensions and the formulation of the
corresponding AdS$_{3}$ duals. In section 3, we study the consistency of the
gravitational bulk under gauge and diffeomorphism invariance and compute the
anomalous currents as well and their algebra. In section 4, we address the
cancellation of anomalies to insure the consistency of the theories. We
provide two cancelling mechanisms: a stringy approach to eliminate the
chiral anomalies and a fermionic mechanism to compensate the gravitational
anomalies caused by the shift between the left and right central charges. In
section 5, by leveraging unitarity constraints, we derive the finiteness
condition given by an upper bound on the rank of the abelian gauge group
which shows that the Landscape of Narain and generalised Narain CFTs in AdS$%
_{3}$ is necessarily finite. We support our results by constructing a
concrete realisation of the finite Narain Landscape by exploiting arguments
from both Swampland and bootstrap Literature. We conclude the paper with a
summary and a discussion of our results.

\section{Narain CFTs, Extensions and AdS$_{3}$ duals}

Narain conformal field theories are undoubtedly among the most well
understood 2D CFTs. Although they were originally developed in the context
of toroidal string compactifications \cite{Het0}, with lattice constructions
\cite{Het1}, they have been recently revisited independently of their string
embedding. Due to their relevance for holographic studies of CFT ensembles
and their averaging, their construction has been extended beyond the usual
even self-dual lattices. Instead, one opts for general lattices with
arbitrary indefinite integral quadratic forms \cite{Meer1, Meer2, Fluc}.
Hereafter, we give a general overview of this recent generalisation.

Standard Narain CFTs are often associated with compactification on
p-dimensional torus $\mathbb{T}^{p}$ where the free bosons are characterised
with momenta valued in an even self dual lattice $\Pi _{p,p}.$\ The Narain
moduli space preserves the T-duality group $O(p,p;\mathbb{Z})$ and is given
by \cite{MW}:%
\begin{equation}
\mathcal{M}_{\Pi _{p,p}}=\frac{O(p,p;\mathbb{Z})\backslash O(p,p;\mathbb{R})%
}{O(p;\mathbb{R})\times O(p;\mathbb{R})}
\end{equation}%
Considering an ensemble of Narain CFTs, the corresponding gravitational dual
can be formulated as an AdS$_{3}$ bulk with $SO(2,2)=SL(2,\mathbb{R}%
)_{L}\times SL(2,\mathbb{R})_{R}$ coupled to $U(1)_{L}^{p-1}\times
U(1)_{R}^{p-1}$ gauge fields \cite{PT}:%
\begin{equation}
\begin{tabular}{lll}
$\mathcal{S}[g_{\mu \nu },A_{\mu },\tilde{A}_{\mu }]$ & $=$ & $\frac{1}{%
16\pi G_{N}}\dint\nolimits_{AdS_{3}}\sqrt{-g}\left( R-2\Lambda \right) $ \\
&  & $+\frac{1}{4\pi }\dsum\limits_{i,j=2}^{p}\dint\nolimits_{AdS_{3}}\left(
A^{i}K_{ij}dA^{j}-\tilde{A}^{i}\tilde{K}_{ij}d\tilde{A}^{j}\right) $%
\end{tabular}%
\end{equation}%
provided one works in the diagonal representation of the gauge fields, where
the boundary gauge fields are expressed in the Cartan basis of the $SL(2,%
\mathbb{R})$ symmetry \cite{SH1}-\cite{SH3}. The $K_{ij}$ is an invertible
symmetric matrix with even integer entries. The diagonal components
represent the level of the individual U(1) gauge fields while the
off-diagonal elements give the mixed terms. The resulting boundary current
algebra arises as $u(1)_{K}^{p}\oplus u(1)_{\tilde{K}}^{p}$ \cite{Fluc, PT}$%
. $

A generalisation of this theory can be achieved if one allows the momenta to
take values beyond the even self dual lattice $\Pi _{p,p},$\ say in an
integral lattice $\Lambda _{p,q}$ equipped with an arbitrary integral
quadratic form \textrm{Q}. The new generalised Narain moduli space is as
follows \cite{Meer1}:
\begin{equation}
\mathcal{M}_{\mathrm{Q}}=\frac{O_{\mathrm{Q}}(\mathbb{Z})\backslash O(p,q;%
\mathbb{R})}{O(p;\mathbb{R})\times O(q;\mathbb{R})}
\end{equation}%
where the set $O_{\mathrm{Q}}(\mathbb{Z})$ is a subgroup of the\textrm{\ }%
orthogonal transformations\textrm{\ }in\textrm{\ }$O\left( p,q,\mathbb{Z}%
\right) $ that preserves the quadratic form $\mathrm{Q}.$ These newly
established models encompass the standard $p=q$ case while also allowing for
$p\neq q,$ which introduces gravitational anomalies. Moreover, they exhibit
exotic modular properties seeing that the underlying lattice is not
necessarily even self dual which affects the modular invariance of the
associated partition function, see \cite{Meer1, Meer2}.

The corresponding bulk structure is equally interesting. Depending on
whether the form $\mathrm{Q}$ is even or odd, different configurations
arise. In the case of an even lattice, one can tune the bulk to be a typical
CS theory diagonally equivalent to an AdS$_{3}$ gravity with $SL(2,\mathbb{R}%
)_{L}\times SL(2,\mathbb{R})_{R}$ symmetry coupled to $U(1)_{L}^{p-1}\times
U(1)_{R}^{q-1}$ gauge fields as follows \cite{Fluc}:%
\begin{equation}
\begin{tabular}{lll}
$\mathcal{S}[g_{\mu \nu },A_{\mu },\tilde{A}_{\mu }]$ & $=$ & $\frac{1}{%
16\pi G_{N}}\dint\nolimits_{AdS_{3}}\sqrt{-g}\left( R-2\Lambda \right) $ \\
&  & $+\frac{1}{4\pi }\dsum\limits_{i,j=1}^{p-1}\dint\nolimits_{AdS_{3}}A^{i}%
\mathrm{Q}_{ij}dA^{j}$ \\
&  & $-\frac{1}{4\pi }\dsum\limits_{a,b=1}^{q-1}\dint\nolimits_{AdS_{3}}%
\tilde{A}^{a}\mathrm{\tilde{Q}}_{ab}d\tilde{A}^{b}$%
\end{tabular}%
\end{equation}%
Otherwise, for an odd lattice, the theory can be formulated as a spin CS
theory defined by the extension on a bounding 4-manifold\textrm{\ }$M_{4}$
with boundary $M_{3}=\partial M_{4}$ such that \cite{Meer1, SpinCS}:%
\begin{equation}
S_{CS}=\frac{i}{8\pi }\int_{M_{3}}A^{\text{\textsc{a}}}\mathrm{Q}_{\text{%
\textsc{ab}}}^{{\small (p,q)}}dA^{\text{\textsc{b}}}=2\pi i\int_{M_{4}}\frac{%
1}{8\pi ^{2}}\frac{dA^{\text{\textsc{a}}}\mathrm{Q}_{\text{\textsc{ab}}}^{%
{\small (p,q)}}dA^{\text{\textsc{b}}}}{2}
\end{equation}%
Note that, regardless of the specific bulk theory one may consider, both
formulations are parameterized by the lattice $\mathrm{Q}$, which dictates
their structure as well as their properties.

\section{Anomalous boundary dynamics}

In this section, we study the emerging anomalies of the AdS$_{3}$ action
coupled to U(1) gauge fields serving as the bulk dual for both Narain and
generalised Narain CFTs. We are interested in two types of anomalies, gauge
and gravitational. We begin by examining the consistency of these models
under gauge and diffeomorphism transformations, and derive the corresponding
anomalous currents. The subsequent sections will address the restoration of
gauge and diffeomorphism invariance.

\subsection{Gauge anomalies in Narain ensemble}

The AdS$_{3}$ gravitational dual to an ensemble of Narain CFTs is given by
the standard AdS$_{3}$ gravity with Lorentz symmetry $SO(2,2)$ coupled to
abelian $U(1)^{2(p-1)}$ gauge fields, provided that diagonal boundary
conditions are imposed \cite{PT, Fluc} as detailed later on. The resulting
action%
\begin{equation}
\begin{tabular}{lll}
$\mathcal{S}[g_{\mu \nu },A_{\mu },\tilde{A}_{\mu }]$ & $=$ & $\mathcal{S}%
_{grav}+\mathcal{S}_{gauge}$ \\
$\mathcal{S}_{grav}[g_{\mu \nu }]$ & $=$ & $\frac{1}{16\pi G_{N}}%
\dint\nolimits_{AdS_{3}}\sqrt{-g}\left( R-2\Lambda \right) $ \\
$\mathcal{S}_{gauge}[A_{\mu },\tilde{A}_{\mu }]$ & $=$ & $\frac{1}{4\pi }%
\dsum\limits_{i,j=2}^{p}\dint\nolimits_{AdS_{3}}\left( A^{i}K_{ij}dA^{j}-%
\tilde{A}^{i}\tilde{K}_{ij}d\tilde{A}^{j}\right) $%
\end{tabular}%
\end{equation}%
is considered below under both gauge and diffeomorphism transformations. We
first study the pure gravitational term $\mathcal{S}_{grav}$ and then move
on to the gauge sector $\mathcal{S}_{gauge}$.

\subsubsection{$\mathcal{S}_{grav}$ as CS gauge theory}

The action $\mathcal{S}_{grav}$ represents the gravitational sector and can
be formulated as a gauged theory, particularly as the difference between two
CS theories each valued in $\left( A^{G},\tilde{A}^{G}\right) \in SL(2,%
\mathbb{R})_{L}\times SL(2,\mathbb{R})_{R}$ \cite{Finit4, GR,AT, W, Y1, Y2}:
\begin{eqnarray}
\mathcal{S}_{grav} &=&\frac{1}{16\pi G_{N}}\dint\nolimits_{AdS_{3}}d^{3}x%
\sqrt{-g}\left( R-2\Lambda \right)  \label{equi} \\
&=&\frac{k^{G}}{4\pi }\int tr(A^{G}dA^{G})-\frac{\tilde{k}^{G}}{4\pi }\int
tr(\tilde{A}^{G}d\tilde{A}^{G})
\end{eqnarray}%
where $k^{G},$ $\tilde{k}^{G}$ are the CS levels related to the 3D Newton
constant $G_{N}$ as $k^{G}=\tilde{k}^{G}=l_{AdS_{3}}/(4G_{N}).$ We have
added the superscript G to denote the CS connections and related parameters
associated with the gravitational sector, allowing for future distinction
once we introduce the gauge fields $\left( A^{g},\tilde{A}^{g}\right) $ of
the abelian gauge sector U(1)$^{2p-2}$. Furthermore, one can also extend the
symmetry $SL(2,\mathbb{R})_{L}\times SL(2,\mathbb{R})_{R}$ to incorporate
higher spin fields by promoting the symmetry to $SL(N,\mathbb{R})_{L}\times
SL(N,\mathbb{R})_{R}$ or to other real forms of higher spin families \cite%
{Satake, BE, BE2}. For the remainder of this discussion however, we will maintain the
gravitational sector in its standard $SL(2,\mathbb{R})$ symmetry.

Focusing on the left sector, and the same analysis applies identically to
the right sector, the infinitesimal gauge change $A^{G}\rightarrow
A^{G}+\delta A^{G}$ gives:
\begin{equation}
\delta \mathcal{S}_{grav}=\frac{k^{G}}{2\pi }\int_{AdS_{3}}tr\left[ \delta
A^{G}\left( dA^{G}+\left( A^{G}\right) ^{2}\right) \right] +\frac{k^{G}}{%
4\pi }\int_{\partial AdS_{3}}tr(\delta A^{G}A^{G})
\end{equation}%
indicating that the action has no extremum on-shell as it doesn't completely
vanish by using the equation of motion $F^{G}=dA^{G}+\left( A^{G}\right)
^{2}=0$. To treat the boundary term, we first reparameterise the boundary of
AdS$_{3}$ with metric $h_{\alpha \beta }\left( z,\bar{z}\right) $ and
complex coordinates $(z,\bar{z}):=\xi $ where $z=\left( \varphi +it\right) /%
\sqrt{2}$ and $\bar{z}=\left( \varphi -it\right) /\sqrt{2}.$ Then, we
rewrite the asymptotic bulk 1-form gauge potentials $A^{G},\tilde{A}^{G}$ as
follows:
\begin{equation}
\begin{tabular}{lllll}
$A_{r}^{G}$ & $\rightarrow $ & $a_{r}=cte$ & $\simeq $ & $\mathrm{g}%
^{-1}\partial _{r}\mathrm{g}$ \\
$A_{\varphi }^{G},A_{t}^{G}$ & $\rightarrow $ & $a_{z},$ $a_{\bar{z}}$ & $%
\simeq $ & $\mathrm{g}^{-1}a_{\xi }\mathrm{g}+\mathrm{g}^{-1}\partial _{\xi }%
\mathrm{g}$%
\end{tabular}
\label{20}
\end{equation}%
and similarly for the right sector's connections. The boundary fields are
diagonal and expand in the diagonal U(1)$^{G}$ of $SL(2,\mathbb{R}%
)_{L}\times SL(2,\mathbb{R})_{R}$ as:%
\begin{equation}
a=\left( a_{z}^{0}dz+a_{\bar{z}}^{0}d\bar{z}\right) K_{0}\qquad ,\qquad
\tilde{a}=\left( \tilde{a}_{z}^{0}dz+\tilde{a}_{\bar{z}}^{0}d\bar{z}\right)
K_{0}  \label{bc}
\end{equation}%
The $SL\left( 2,\mathbb{R}\right) $ group elements $\mathrm{g}$ and $\mathrm{%
\tilde{g}}$ of the gauge transformation (\ref{20}) are set in the
Grumiller-Riegler (GR) gauge as \cite{GR}:
\begin{equation}
\mathrm{g}=e^{rK_{0}}e^{K_{-}},\qquad \mathrm{g}^{-1}\frac{\partial \mathrm{g%
}}{\partial r}=K_{0},\qquad \mathrm{\tilde{g}=g}^{-1}  \label{radial2}
\end{equation}%
with $SL\left( 2,\mathbb{R}\right) $ generators $K_{0},$ $K_{\pm }$ obeying%
\begin{equation}
\left[ K_{n},K_{m}\right] =\left( n-m\right) K_{n+m},\qquad n,m=0,\pm
\label{k}
\end{equation}%
Consequently, the bulk 1-form gauge potential $A^{G}=A_{t}^{G}dt+A_{\varphi
}^{G}d\varphi ,$ at the boundary, is valued in the diagonal U(1)$^{G}$ and
expand in the canonical 1-forms like $a_{z}dz+a_{\bar{z}}d\bar{z}$. The
variation becomes:%
\begin{equation}
\delta \mathcal{S}_{grav}=\frac{k^{G}}{4\pi }\int_{\partial AdS_{3}}d^{2}z%
\sqrt{\left\vert \mathrm{h}\right\vert }\left( \delta a_{z}a_{\bar{z}%
}-\delta a_{\bar{z}}a_{z}\right)
\end{equation}%
with $\mathrm{h}$ the determinant of the boundary metric $h_{\alpha \beta
}\left( z,\bar{z}\right) $. In an attempt to cancel this deformation, we
introduce the boundary term $\mathcal{S}_{grav}^{bnd}$ given by:%
\begin{equation}
\mathcal{S}_{grav}^{bnd}=\frac{k^{G\prime }}{4\pi }\int\nolimits_{\partial
AdS_{3}}d^{2}z\sqrt{\left\vert \mathrm{h}\right\vert }(a_{z}a_{\bar{z}})
\end{equation}%
For $k^{G\prime }=k^{G},$ the total varied action for the left sector is:%
\begin{equation}
\delta \mathcal{S}_{grav}^{tot}=\delta (\mathcal{S}_{grav}+\mathcal{S}%
_{grav}^{bnd})=\frac{k^{G}}{2\pi }\int\nolimits_{\partial AdS_{3}}d^{2}z%
\sqrt{\left\vert \mathrm{h}\right\vert }\left[ \delta a_{z}a_{\bar{z}}\right]
\label{bndy}
\end{equation}%
A good variational principle is therefore restored by taking $\delta a_{z}=0$
( while $a_{\bar{z}}$ free) which confirms with Grumiller-Riegler boundary
conditions preventing any gauge anomalies \cite{GR}$.$ This boundary
constraint leaves $a_{\bar{z}}$ as the physical current. Combined with the
complementary right sector, which features the current $\tilde{a}_{z}$, the
resulting theory exhibits two copies of the affine Kac-Moody (KM) algebra $%
u(1)_{k^{G}}\times u(1)_{\tilde{k}^{G}}$ \cite{GR, GR2}:
\begin{equation}
\left\{ a_{\bar{z}}{\small (\xi )},a_{\bar{z}}{\small (\xi }^{\prime }%
{\small )}\right\} _{DB}=\frac{k^{G}}{2\pi }\frac{\partial }{\partial z}%
\delta _{2}\left( {\small \xi },{\small \xi }^{\prime }\right)
\end{equation}%
and%
\begin{equation}
\left\{ \tilde{a}_{z}{\small (\xi )},\tilde{a}_{z}{\small (\xi }^{\prime }%
{\small )}\right\} _{DB}=-\frac{\tilde{k}^{G}}{2\pi }\frac{\partial }{%
\partial \bar{z}}\delta _{2}\left( {\small \xi },{\small \xi }^{\prime
}\right)
\end{equation}%
having the central charges $c=\tilde{c}=3l_{AdS_{3}}/(2G_{N}).$

Next, regarding the gravitational anomalies of this theory, it is anomaly
free due to the fact that the central charges of the left and right sectors
are equal: $c=\tilde{c}$. Otherwise, we must take on the non-vanishing
gravitational CS term \cite{kraus1}:%
\begin{equation}
\mathcal{S}_{grav}^{CS}\left( \Gamma \right) =-\frac{c-\tilde{c}}{96\pi }%
\int_{AdS_{3}}Tr\left( \Gamma d\Gamma +\frac{2}{3}\Gamma ^{3}\right)
\end{equation}%
where the divergence of the stress tensor $T^{\mu \nu }$ follows from the
breakdown of invariance under the diffeomorphisms $\Gamma \rightarrow \Gamma
+\delta \Gamma ;$ it is given by:%
\begin{equation}
\partial _{\mu }T^{\mu \nu }=-\frac{c-\tilde{c}}{96\pi }g^{\nu \alpha
}\epsilon ^{\mu \rho }\partial _{\mu }\partial _{\beta }\Gamma _{\alpha \rho
}^{\beta }
\end{equation}%
The shift in the central charges should be of the same magnitude but with
opposite signs \cite{kraus2}, therefore:%
\begin{equation}
\begin{tabular}{ccc}
$c-\tilde{c}$ & $=$ & $96\pi \beta ,$ \\
$c$ & $=$ & $\frac{3l_{AdS_{3}}}{2G}+48\pi \beta ,$ \\
$\tilde{c}$ & $=$ & $\frac{3l_{AdS_{3}}}{2G}-48\pi \beta .$%
\end{tabular}%
\end{equation}%
We will however proceed with the first case of equal central charges and
disregard the gravitational CS term.

Before moving forward with the abelian gauge sector, we should bear in mind
that the non chiral gravitational sector is free from both gravitational and
U(1)$^{G}$ gauge anomalies. Any possible emerging anomalies can, a priori,
only arise from the abelian U(1)$^{2p-2}$ gauge sector which we will work on
next. We summarize this information in table \ref{tab1}%
\begin{table}[H]
\centering
\begin{tabular}{|c||c|c|}
\hline
{\small Bulk theory} & \multicolumn{2}{||c|}{{\small Standard Narain (}$%
{\small p=q}${\small )}} \\ \hline\hline
{\small Anomalies} & {\small CS U(1)}$^{G}${\small \ sector} & {\small %
Abelian U(1)}$^{2p-2}${\small \ sector} \\ \hline
{\small Gravitational} & {\small
\begin{tabular}{c}
no \\
$c=\tilde{c}$%
\end{tabular}%
} & {\small ?} \\ \hline
{\small Gauge} & {\small
\begin{tabular}{c}
no \\
$\delta a_{z}=0$%
\end{tabular}%
} & {\small ?} \\ \hline
\end{tabular}%
    \caption{Anomaly structure in the bulk dual of the standard Narain ensemble. Here, "no" indicates that the theory is anomaly free provided that $c=\tilde{%
c}$\ and $\delta a_{z}=0$.}
\label{tab1}
\end{table}

\subsubsection{Anomalies of the abelian gauge sector $\mathcal{S}_{gauge}$}

Here, we proceed to the abelian gauge sector which is given by the action $%
\mathcal{S}_{gauge}$ expressed as:%
\begin{equation}
\mathcal{S}_{gauge}=\frac{1}{4\pi }\dsum\limits_{i,j=2}^{p}\dint%
\nolimits_{AdS_{3}}\left( A^{g_{i}}K_{ij}dA^{g_{j}}-\tilde{A}^{g_{i}}\tilde{K%
}_{ij}d\tilde{A}^{g_{j}}\right)
\end{equation}%
we use the superscript $g$ to denote the gauge CS quantities to distinguish
them from $G$ labeling the gravitational sector and subsequently omit the
matrix $K_{ij}$ for simpler notations. This action, like the gravitational
one, consists of the difference between two CS terms, but now with
connections valued in the abelian $U(1)_{L}^{p-1}\times U(1)_{R}^{p-1}$
instead of $SL\left( 2,\mathbb{R}\right) _{L}\times SL\left( 2,\mathbb{R}%
\right) _{R}.$ Following the previous analysis, we introduce the boundary
term for the left half with symmetry $U(1)_{L}^{p-1}$:%
\begin{equation}
\mathcal{S}_{gauge}^{bnd}=\frac{1}{4\pi }\dint\nolimits_{\partial
AdS_{3}}d^{2}z\left( A_{z}^{g}A_{\bar{z}}^{g}\right)
\end{equation}%
Under an infinitesimal gauge change $A^{g}\rightarrow A^{g}+\delta A^{g}$,
the variation of the action for the left sector leaves the boundary term:
\begin{equation}
\delta \left( \mathcal{S}_{gauge}+\mathcal{S}_{gauge}^{bnd}\right) =\frac{1}{%
2\pi }\dint\nolimits_{\partial AdS_{3}}\delta A_{z}^{g}A_{\bar{z}}^{g}
\end{equation}%
At the asymptotic boundary of AdS$_{3},$ the gauge field expands as \cite%
{kraus1, S1, S2}:%
\begin{equation}
A^{g}\left( r,\xi \right) \sim A_{\left( 0\right) }^{g}+r^{1-\Delta
}A_{\left( 2\right) }^{g}+...
\end{equation}%
where the leading term $A_{\left( 0\right) }^{g}$ defines the non dynamical
background of the gauge field; it is flat since it doesn't decay with $r$.
The subleading term, $r^{1-\Delta }A_{\left( 2\right) }^{g},$\ represents
the normalisable mode of $A^{g},$ with the exponent $1-\Delta <0$ indicating
the duality of $A^{g}$ to a vector primary operator of dimension $\Delta .$
At the boundary, as $r\rightarrow \infty $, the field $A^{g}$ falls off to $%
A_{\left( 0\right) }^{g}.$ Consequently, the variation becomes for the left
sector:%
\begin{equation}
\delta \left( \mathcal{S}_{gauge}+\mathcal{S}_{gauge}^{bnd}\right) =\frac{1}{%
2\pi }\dint\nolimits_{\partial AdS_{3}}\delta \left( A_{z}^{g}\right)
_{\left( 0\right) }\left( A_{\bar{z}}^{g}\right) _{\left( 0\right)
}=\int\nolimits_{\partial AdS_{3}}d^{2}z\sqrt{\left\vert \mathrm{h}%
\right\vert }\mathcal{J}_{\bar{z}}\delta \left( A_{z}^{g}\right) _{\left(
0\right) }
\end{equation}%
To ease notations, we will drop the 0 subscript. The associated chiral
current $\mathcal{J}_{\bar{z}}$ is given by:%
\begin{equation}
\mathcal{J}_{\bar{z}i}=\frac{1}{2\pi }K_{ij}A_{\bar{z}}^{g_{j}}
\end{equation}%
Extending this analysis to the full gauge sector with an enhanced $%
U(1)_{L}^{p-1}\times U(1)_{R}^{p-1}$ while considering CS matrix $%
K_{ij}^{\prime }=K_{ij}$ for the left boundary gauge fields $A_{z}$ and CS
matrix $\tilde{K}_{ij}^{\prime }=-\tilde{K}_{ij}$ for $\tilde{A}_{\bar{z}},$
we obtain:%
\begin{equation}
\delta \left( \mathcal{S}_{gauge}+\mathcal{S}_{gauge}^{bnd}\right) =\frac{1}{%
2\pi }\int\nolimits_{\partial AdS_{3}}\delta A_{z}^{g}A_{\bar{z}}^{g}-\frac{1%
}{2\pi }\int\nolimits_{\partial AdS_{3}}\delta \tilde{A}_{\bar{z}}^{g}\tilde{%
A}_{z}^{g}
\end{equation}%
The associated chiral currents are:
\begin{equation}
\mathcal{J}_{\bar{z}i}=\frac{1}{2\pi }K_{ij}A_{\bar{z}}^{g_{j}}\qquad
,\qquad \mathcal{\tilde{J}}_{zi}=-\frac{1}{2\pi }\tilde{K}_{ij}\tilde{A}%
_{z}^{g_{j}}  \label{anoc}
\end{equation}%
with Laurent modes as follows:%
\begin{equation}
J_{n}^{i}=\oint \frac{d\bar{z}}{2i\pi }\bar{z}^{n-1}\mathcal{J}_{\bar{z}%
}^{i}\qquad ,\qquad \tilde{J}_{n}^{i}=-\oint \frac{dz}{2i\pi }z^{n-1}%
\mathcal{\tilde{J}}_{z}^{i}
\end{equation}%
where $i,j=2,...,p.$ The corresponding abelian affine algebra is:%
\begin{equation}
\left[ J_{m}^{i},J_{n}^{j}\right] =\frac{1}{2}mK^{ij}\delta _{m+n},_{0}
\label{ex2}
\end{equation}%
with central charge $c_{g}=rank[U(1)^{p-1}]=p-1$ for the left sector. And%
\begin{equation}
\left[ \tilde{J}_{m}^{i},\tilde{J}_{n}^{j}\right] =-\frac{1}{2}m\tilde{K}%
^{ij}\delta _{m+n},_{0}
\end{equation}%
having $\tilde{c}_{g}=p-1$ for the right sector.

In general, the 2D chiral anomaly represents the violation of the
conservation of chiral currents as follows \cite{S1,S2}:%
\begin{equation}
\nabla ^{\mu }\mathcal{J}_{\mu a}=\frac{q_{ab}}{8\pi }F_{\mu \nu
}^{b}\varepsilon ^{\mu \nu }  \label{L1}
\end{equation}%
where $q_{ab}$ are 't Hooft anomaly coefficients. This can be related to the
currents (\ref{anoc}) as their divergences are given by:%
\begin{equation}
\nabla _{\bar{z}}\mathcal{J}_{\bar{z}i}=\frac{1}{2\pi }K_{ij}\nabla _{\bar{z}%
}A_{\bar{z}}^{g_{j}}\qquad ,\qquad \nabla _{z}\mathcal{\tilde{J}}_{zi}=-%
\frac{1}{2\pi }\tilde{K}_{ij}\nabla _{z}\tilde{A}_{z}^{g_{j}}
\end{equation}

As for the gravitational anomalies, none arise in this sector since the
central charges are equal; $c_{g}=\tilde{c}_{g}=p-1$.\newline
In conclusion, the gauge sector associated with an ensemble of Narain CFTs
having equal left and right central charges gives rise to 2D chiral
anomalies and extend the current algebra of the gravitational sector U(1)$%
^{G}$ by additional copies of the U(1)$^{p-1}\times $U(1)$^{p-1}$ current
algebras. The sector is however free of any gravitational anomalies.%
\begin{table}[H]
\centering
\begin{tabular}{|c||c|c|}
\hline
{\small Bulk theory} & \multicolumn{2}{||c|}{\small Standard Narain (p=q)}
\\ \hline\hline
{\small Anomalies} & {\small CS U(1)}$^{G}${\small \ sector\ } & {\small %
Abelian U(1)}$^{2p-2}${\small \ sector} \\ \hline
{\small Gravitational} & {\small
\begin{tabular}{c}
no \\
$c=\tilde{c}$%
\end{tabular}%
} & {\small
\begin{tabular}{c}
no \\
$c_{g}=\tilde{c}_{g}$%
\end{tabular}%
} \\ \hline
{\small Gauge} & {\small
\begin{tabular}{c}
no \\
$\delta a_{z}=0$%
\end{tabular}%
} & {\small
\begin{tabular}{c}
yes \\
$\nabla _{\bar{z}}\mathcal{J}_{\bar{z}i}\neq 0,$ $\nabla _{z}\mathcal{\tilde{%
J}}_{zi}\neq 0$%
\end{tabular}%
} \\ \hline
\end{tabular}%
\caption{The gauge sector associated with an ensemble of Narain CFTs
having equal left and right central charges gives rise to 2D chiral
anomalies but remains however free of any gravitational ones.}
\label{tab12}
\end{table}

\subsection{Generalised NCFTs}

For generalised Narain CFTs, the analysis is largely similar to the standard
case with a few notable differences. If we confine the study to the case of
an even lattice $\mathrm{Q,}$ the bulk gravitational action retains the same
structure as that of Narain CFTs except that the number of added U(1) gauge
fields differs between left and right sectors. More specifically, we
introduce $p$ U(1) gauge field in the left moving sector; and $q$ ones in
the right moving one. The resulting bulk action is:%
\begin{equation}
\begin{tabular}{lll}
$\mathcal{S}[g_{\mu \nu },A_{\mu },\tilde{A}_{\mu }]$ & $=$ & $\frac{1}{%
16\pi G_{N}}\dint\nolimits_{AdS_{3}}\sqrt{-g}\left( R-2\Lambda \right) $ \\
&  & $+\frac{1}{4\pi }\dsum\limits_{a,b=1}^{p-1}\dint\nolimits_{AdS_{3}}A^{a}%
\mathrm{Q}_{ab}dA^{b}$ \\
&  & $-\frac{1}{4\pi }\dsum\limits_{c,d=1}^{q-1}\dint\nolimits_{AdS_{3}}%
\tilde{A}^{c}\mathrm{\tilde{Q}}_{cd}d\tilde{A}^{d}$%
\end{tabular}
\label{act2}
\end{equation}

A number of consequences follow from this left and right asymmetry. For
instance, while the usual moduli space is defined in terms of the background
metric G and two-form B, in the generalised case it is parameterised by the
indefinite non-unimodular lattices $\mathrm{Q}_{ab}$ and $\mathrm{\tilde{Q}}%
_{cd}.$ In addition, the existence of a modular invariant partition function
is no longer guaranteed. However, our primary interest lies in analyzing the
anomalies of the bulk theories, so we will not elaborate further on the
other differences; report to \cite{Fluc, M3} for details.

Regarding the gravitation part of the action (\ref{act2}), it remains
unchanged and is still free of both gauge anomalies and gravitational ones
provided that certain conditions are applied. In particular, we must
distinguish between the different central charges within the theory. The
gravitational central charges, $c=3l_{AdS_{3}}/2G_{N}$ and $\tilde{c}%
=3l_{AdS_{3}}/2G_{N}$, associated with the left and right moving
gravitational sectors under the $SL(2,R)_{L}\times SL(2,R)_{R}$ symmetry are
assumed to be equal here. Otherwise, one would require to include the
gravitational Chern-Simons term to exhibit the breaking of non-chirality.
Furthermore, there are other central charges, $c_{g}=p-1$ and $\tilde{c}%
_{g}=q-1$, associated with the current algebras arising from the gauge
sector; they are not equal in the generalised case which signals the
existence of a gravitational anomaly. Since this anomaly is sourced by
abelian gauge terms, and to avoid any confusion, we refer to it as
gauged-sourced gravitational anomaly. Its cancellation will be discussed in
more details shortly after. For now, we emphasize that the pure
gravitational central charges are equal, which ensures the absence of pure
gravitational anomalies. On the other hand, a gauged-sourced gravitational
anomaly does emerge due to the asymmetry in the number of added left and
right U(1) currents.

Now turning to gauge anomalies, their structure follows the same previous
computation as for Narain theories with anomalous currents
\begin{equation}
\nabla _{\bar{z}}\mathcal{J}_{\bar{z}a}=\frac{1}{2\pi }\mathrm{Q}_{ab}\nabla
_{\bar{z}}A_{\bar{z}}^{g_{b}}\qquad ,\qquad \nabla _{z}\mathcal{\tilde{J}}%
_{zc}=-\frac{1}{2\pi }\mathrm{\tilde{Q}}_{cd}\nabla _{z}\tilde{A}_{z}^{g_{d}}
\end{equation}%
and associated algebras taking the form:%
\begin{equation}
\left[ J_{m}^{a},J_{n}^{b}\right] =\frac{1}{2}m\mathrm{Q}^{ab}\delta
_{m+n},_{0}\qquad ,\qquad \left[ \tilde{J}_{m}^{c},\tilde{J}_{n}^{d}\right]
=-\frac{1}{2}m\mathrm{\tilde{Q}}^{cd}\delta _{m+n},_{0}
\end{equation}%
with central charges $c_{g}=p-1$ and $\tilde{c}_{g}=q-1$. The indices run
over $a,b=1...p-1$ and $c,d=1...q-1$ for the $p$ left and $q$ right U(1)
gauge fields respectively.

\section{Restoring consistency via anomaly cancellation}

In this section, we investigate the restoration of the consistency of the
theories by cancelling the anomalies. To begin, we briefly summarize the
various types of anomalies previously derived, before addressing their
cancellation mechanisms:%
\begin{table}[H]
\centering
\begin{tabular}{|c||c|c||c|c|}
\hline
{\small Bulk theory} & \multicolumn{2}{||c||}{\small Standard Narain (p=q)}
& \multicolumn{2}{||c|}{{\small Generalized Narain (p}${\small \neq }$%
{\small q)}} \\ \hline\hline
{\small Anomalies} & {\small U(1)}$^{G}${\small \ sector} & {\small U(1)}$%
^{2p-2}${\small \ sector} & {\small U(1)}$^{G}${\small \ sector} & {\small %
U(1)}$^{p+q-2}$ {\small sector} \\ \hline
{\small Gravitational} & {\small
\begin{tabular}{c}
no \\
$c=\tilde{c}$%
\end{tabular}%
} & {\small
\begin{tabular}{c}
no \\
$c_{g}=\tilde{c}_{g}$%
\end{tabular}%
} & {\small
\begin{tabular}{c}
no \\
$c=\tilde{c}$%
\end{tabular}%
} & {\small
\begin{tabular}{c}
yes \\
$c_{g}\neq \tilde{c}_{g}$%
\end{tabular}%
} \\ \hline
{\small Gauge} & {\small
\begin{tabular}{c}
no \\
$\delta a_{z}=0$%
\end{tabular}%
} & {\small
\begin{tabular}{c}
yes \\
$\nabla _{\bar{z}}\mathcal{J}_{\bar{z}i}\neq 0$ \\
$\nabla _{z}\mathcal{\tilde{J}}_{zi}\neq 0$%
\end{tabular}%
} & {\small
\begin{tabular}{c}
no \\
$\delta a_{z}=0$%
\end{tabular}%
} & {\small
\begin{tabular}{c}
yes \\
$\nabla _{\bar{z}}\mathcal{J}_{\bar{z}i}\neq 0$ \\
$\nabla _{z}\mathcal{\tilde{J}}_{zi}\neq 0$%
\end{tabular}%
} \\ \hline
\end{tabular}
\caption{ The gravitational sector is anomaly
free for both Narain and generalised CFTs whereas the gauge sector exhibits anomalies for both settings.}
\label{res}
\end{table}%
From the table, it is clear that the gravitational sector remains anomaly
free for both Narain and generalised CFTs and does not require further
attention. In contrast, the gauge sector exhibits anomalies in both
settings. Therefore, we subsequently introduce two distinct mechanisms: one
to cancel the gauge anomalies for both types of CFTs; and another to
eliminate the gauged-sourced gravitational anomalies arising in the
generalised case.

\subsection{Stringy resolution of the chiral anomaly}

To guarantee the consistency of the EFT, both gauge and gravitational
anomalies at the boundary must be canceled. To start, we focus on the gauge
anomalies and defer the gravitational anomalies to the following subsection.

The gauge anomalies equivalent to a 2D chiral anomaly manifest as a
violation of the conservation of chiral currents:
\begin{equation}
\nabla _{\bar{z}}\mathcal{J}_{\bar{z}i}=\frac{1}{2\pi }K_{ij}\nabla _{\bar{z}%
}A_{\bar{z}}^{g_{j}}\qquad ,\qquad \nabla _{z}\mathcal{\tilde{J}}_{zi}=-%
\frac{1}{2\pi }\tilde{K}_{ij}\nabla _{z}\tilde{A}_{z}^{g_{j}}
\end{equation}%
In conventional string theory models, such anomalies usually get typically
canceled by contributions from the worldsheet degrees of freedom as in \cite%
{C5, Finit2}. Building on this idea, a similar mechanism has been proposed
in \cite{Finit4} for higher spin gravity theories in AdS$_{3}$ and will be
revisited here and adapted to our setting. The key idea is to exploit the
duality between Chern-Simons and WZW actions by cancelling the CS anomalies
by introducing compensating terms at the boundary via a string defined by a
non chiral WZW action.

The difference between the two CS actions, $\mathcal{S}_{{\small CS}}^{%
{\small G}_{L}}-\mathcal{S}_{{\small CS}}^{{\small G}_{R}},$ can be recast
as a difference between two chiral WZW\ actions, $\mathcal{S}_{{\small WZW}%
}^{{\small G}_{L}}\left[ g_{L}\right] -\mathcal{S}_{{\small WZW}}^{{\small G}%
_{R}}\left[ g_{R}\right] $ which defines a non chiral WZW action $\mathcal{S}%
_{WZW}^{NC}\left[ g\right] $ with $g=g_{L}^{-1}g_{R}.$ This latter
decomposes into two blocks:%
\begin{equation}
\mathcal{S}_{WZW}^{NC}\left[ g\right] =\mathcal{S}_{bulk}\left[ g\right] +%
\mathcal{S}_{edge}\left[ g\right]   \label{NCWZW}
\end{equation}%
with bulk term%
\begin{equation}
\mathcal{S}_{bulk}\left[ g\right] =\frac{k^{g}}{12\pi }\int_{AdS_{3}}Tr%
\left( g^{-1}dg\right) ^{3}
\end{equation}%
and boundary term having the typical form:%
\begin{equation}
\mathcal{S}_{edge}\left[ g\right] =\frac{k^{g}}{4\pi }\int_{\partial
AdS_{3}}Tr\left[ \left( g^{-1}\nabla _{z}g\right) \left( g^{-1}\nabla _{\bar{%
z}}g\right) \right]
\end{equation}%
This non chiral WZW action naturally introduces a 2D CFT interpreted as a
bosonic string on the boundary of our AdS$_{3}.$ In fact, with a suitable
reparameterisation of $g=e^{T_{\text{\textsc{a}}}X^{\text{\textsc{a}}}}$
where $X^{\text{\textsc{a}}}$ are scalar fields and $T_{\text{\textsc{a}}}$
are generators of the gauge algebra with metric $\mathrm{G}_{\text{\textsc{ab%
}}}=\left\langle T_{\text{\textsc{a}}},T_{\text{\textsc{b}}}\right\rangle $.
The boundary action becomes:%
\begin{equation}
\int_{\partial AdS_{3}}d^{2}\xi \sqrt{|-h|}h^{\alpha \beta }\mathrm{G}_{%
\text{\textsc{ab}}}\partial _{\alpha }X^{\text{\textsc{a}}}\partial _{\beta
}X^{\text{\textsc{b}}}
\end{equation}%
where $h_{\alpha \beta }$ is the metric of the boundary of AdS$_{3}$. This
has the standard form of a free bosonic string in the conformal gauge on the
group manifold G with metric $\mathrm{G}_{\text{\textsc{ab}}}$ \cite{wzw3}.

Using the decomposition $g=g_{L}^{-1}g_{R}$ \cite{wbh,wzw3}, the non-chiral
WZW can be split into two chiral WZW actions: a left block with field action
$\mathcal{S}_{WZW}^{L}\left[ g_{L}\right] $; and a right one with field
action $\mathcal{S}_{WZW}^{R}\left[ g_{R}\right] .$ The coupling of the left
WZW action to the background gauge field $A_{z}$ reads as \cite{S1,S2}:%
\begin{eqnarray}
\mathcal{S}_{WZW}^{L}\left( g_{L}\right)  &=&\frac{1}{4\pi }\int_{\partial
AdS_{3}}d^{2}zTr\left[ \left( g_{L}^{-1}\partial _{z}g_{L}^{-1}\right)
\left( g_{L}\partial _{\bar{z}}g_{L}\right) +2g_{L}^{-1}\partial _{\bar{z}%
}g_{L}A_{z}\right]   \notag \\
&&+\frac{1}{12\pi }\int_{AdS_{3}}Tr\left( g_{L}^{-1}dg_{L}\right) ^{3},
\end{eqnarray}%
and the coupling of the right WZW action to $\tilde{A}_{\bar{z}}$ is given
by a similar term as follows:%
\begin{eqnarray}
\mathcal{S}_{WZW}^{R}\left( g_{R}\right)  &=&\frac{1}{4\pi }\int_{\partial
AdS_{3}}d^{2}zTr\left( g_{R}^{-1}\partial _{z}g_{R}^{-1}g_{R}\partial _{\bar{%
z}}g_{R}-2g_{R}^{-1}\partial _{z}g_{R}\tilde{A}_{\bar{z}}\right) +  \notag \\
&&\frac{1}{12\pi }\int_{AdS_{3}}Tr\left( g_{R}^{-1}dg_{R}\right) ^{3}
\end{eqnarray}%
The above chiral field actions give two currents: a right $\mathcal{J}_{\bar{%
z}}$ and left $\mathcal{\tilde{J}}_{z}$ given by:%
\begin{equation}
\begin{tabular}{lllll}
$\mathcal{J}_{\bar{z}}$ & $=$ & $\frac{1}{2\pi }g_{L}^{-1}\partial _{\bar{z}%
}g_{L}$ & $=$ & $+\frac{1}{2\pi }A_{\bar{z}},$ \\
$\mathcal{\tilde{J}}_{z}$ & $=$ & $-\frac{1}{2\pi }g_{R}^{-1}\partial
_{z}g_{R}$ & $=$ & $-\frac{1}{2\pi }\tilde{A}_{z},$%
\end{tabular}%
\end{equation}%
These currents reproduce the required anomalous structure, but with opposite
signs, precisely as required for cancellation. This mechanism applies to
both Narain and generalised Narain CFTs; the main distinction lies in the
generalised case where the number of right currents is different than that
of the left moving ones.

\subsection{Fermionic compensation of gravitational anomalies}

Generalised Narain theories, in addition to them having gauge anomalies,
carry gauged sourced gravitational anomalies due to the mismatch $\Delta
=\left\vert p-q\right\vert $ which is a consequence of the unequal number of
gauge fields introduced in the left- and right-moving sectors. This is an
anomalous defect that has to be addressed to insure the consistency of the
model. A proposal is to consider adding fermionic degrees of freedom so that
the difference in gauge central charges vanishes and reinstate $\Delta
_{tot}=0$. In heterotic string theory \cite{Ht1, Ht2, Ht3}, this is achieved
by introducing right-moving fermions on the worldsheet which enables to
construct the internal symmetry SO$(32)$ or E$_{8}\times $ E$_{8}$ and
ensures the consistency of the worldsheet. We adapt this mechanism within
our specific lattice-based code CFT framework to balance the central charge,
ensure conformal invariance, and cancels the worldsheet gravitational
anomaly.

Here, we take the gauge boundary potential as a composite object of a 2D CFT
realized in terms of both scalar and chiral fermi fields. We give $A_{\bar{z}%
}^{i}$\ and $\tilde{A}_{z}^{\bar{\imath}}$\ realisations in terms of scalars
$\left( X^{i},\bar{Y}^{\bar{\imath}}\right) $\ and chiral fermions\textrm{\ }%
($\psi _{L},\bar{\chi}_{R}$). As illustration, we consider the case $p<q$
and set $q=p+2n$ where the parameter n is a positive integer that denotes
the number of world sheet fermions. For an even integer lattice, the total
number of fermions is 2n such that $2\leq q\leq 2n\leq p.$

the gauge fields then splits into:
\begin{equation}
A_{\bar{z}}=\left( A_{\bar{z}}\right) _{scalar}+\left( A_{\bar{z}}\right)
_{fermi}\qquad ,\qquad \tilde{A}_{z}=\left( \tilde{A}_{z}\right) _{scalar}
\end{equation}%
with the realisations%
\begin{equation}
\begin{tabular}{lllll}
$\left( A_{\bar{z}}^{i}\right) _{scalar}$ & $=$ & $\frac{\partial }{\partial
\bar{z}}X^{i}$ & $\qquad ,\qquad $ & $1\leq i\leq p$ \\
$\left( A_{\bar{z}}^{\theta }\right) _{fermi}$ & $=$ & $\bar{\psi}_{\bar{z}%
/2}^{\theta }\psi _{\bar{z}/2}^{\theta }$ & $\qquad ,\qquad $ & $1\leq
\theta \leq 2n$ \\
$\left( \tilde{A}_{z}^{\bar{\imath}}\right) _{scalar}$ & $=$ & $\frac{%
\partial }{\partial z}\bar{Y}^{\bar{\imath}}$ & $\qquad ,\qquad $ & $1\leq
\bar{\imath}\leq q+2n$%
\end{tabular}
\label{th}
\end{equation}%
and free field equations of $X^{i},$ $\psi _{\bar{z}/2}^{\theta },\bar{\psi}%
_{\bar{z}/2}^{\theta }$\ and $\bar{Y}^{\bar{\imath}}:$%
\begin{equation}
\frac{\partial }{\partial \bar{z}}\frac{\partial }{\partial z}X^{i}=0\qquad
,\qquad \frac{\partial }{\partial \bar{z}}\psi _{\bar{z}/2}^{\theta
}=0\qquad ,\qquad \frac{\partial }{\partial z}\frac{\partial }{\partial \bar{%
z}}\bar{Y}^{\bar{\imath}}=0
\end{equation}%
By construction, the added chiral fermions adjust the central charge of the
left-moving sector, thereby ensuring that the gauged-sourced gravitational
anomaly is cancelled.

\section{Finiteness of the Narain CFTs Landscape}

\qquad In the AdS$_{3}$ bulk, the AdS distance conjecture expects the
emergence of an infinite tower of massless states upon taking $\Lambda
=-1/l_{AdS_{3}}^{2}\rightarrow 0$ when $l_{AdS_{3}}\rightarrow \infty $
seeing that $m\sim $ $|\Lambda |^{\alpha },$ with $\Lambda $ being the
cosmological constant of the AdS$_{3}$ space and $\alpha $\ a positive
constant that, based on dimensional arguments, is equal to 1/2 in this
instance \cite{DC}.

Additionally, one must also insure the abidance of the finiteness conjecture%
\textrm{\ }\cite{Finit1}-\cite{Finit4}, stipulating the existence of a
finite number of massless modes. For AdS theories, the finiteness constraint
has a sharper formulation where one must impose a cut-off on AdS theories as
follows \cite{rev}:
\begin{equation}
\Lambda _{cut-off}\leq m\sim |\Lambda |^{1/2}\qquad ,\qquad |\Lambda |^{1/2}=%
\frac{1}{l_{AdS_{3}}}
\end{equation}%
where $l_{AdS_{3}}$ is fixed. Accordingly, there are only a finite number of
AdS theories\ with cutoff $\Lambda _{cut-off}$ corresponding to a finite
number of dual CFTs. In fact, using $c=3l_{AdS_{3}}/2G_{N}$ we have $\Lambda
^{1/2}=\frac{3}{2G_{N}c}.$ Thus:
\begin{equation}
c\leq \frac{3}{2G_{N}}\Lambda _{cut-off}^{-1}
\end{equation}%
Additionally, given a unitary 2D boundary CFT of AdS$_{3}$ gravity with left
central charge $c>0;$ and a Kac-Moody algebra $G$ of level \textrm{k}
realized on the left moving sector of the CFT$_{2}$ with central charge as:%
\begin{equation}
0<c_{G}=\frac{k_{L}^{G}\text{ }\dim G}{k_{L}^{G}\text{ }+h^{\vee }\text{ }}
\end{equation}%
where $h_{L}^{\vee }$ is the dual Coxeter number of $G_{L}$, the following
consistency constraint relation is required for the unitarity of the CFT:
\begin{equation}
0<c_{G}<c  \label{ccL}
\end{equation}%
Combining both unitarity and finiteness constraint, we get:%
\begin{equation}
c_{G}<c\leq \frac{3}{2G_{N}}\Lambda _{cut-off}^{-1}
\end{equation}%
For abelian gauge groups, it reads%
\begin{equation}
rank\left( G\right) <c\leq \frac{3}{2G_{N}}\Lambda _{cut-off}^{-1}  \label{F}
\end{equation}%
In the left sector, it translates to:%
\begin{equation}
1\leq p-1<c\leq \frac{3}{2G_{N}}\Lambda _{cut-off}^{-1}  \label{F1}
\end{equation}%
and to%
\begin{equation}
1\leq q-1<c\leq \frac{3}{2G_{N}}\Lambda _{cut-off}^{-1}  \label{F2}
\end{equation}%
for the right one.

The conditions (\ref{F1}) and (\ref{F2}) derived here impose an explicit
upper bound on the number of abelian gauge symmetries U(1) that one can
couple to AdS$_{3}$ gravity. In consequence, this also bounds the number of
admissible Narain CFTs composing the ensemble that describes the holographic
dual. Particularly, (\ref{F1}) and (\ref{F2}) clearly delineate the finite
region of the Landscape resided by Narain CFTs. This proves that the
presence of the AdS$_{3}$ dual tightly constrains the ensemble of Narain
CFTs in accordance with Swampland criteria. In the graph \textbf{\ref{finit}}%
, per illustration, we capture the bounded region representing the Narain
CFT Landscape for the left sector constrained by both the finiteness
conjecture and the unitarity constraints:
\begin{figure}[h]
\begin{center}
\includegraphics[width=17cm]{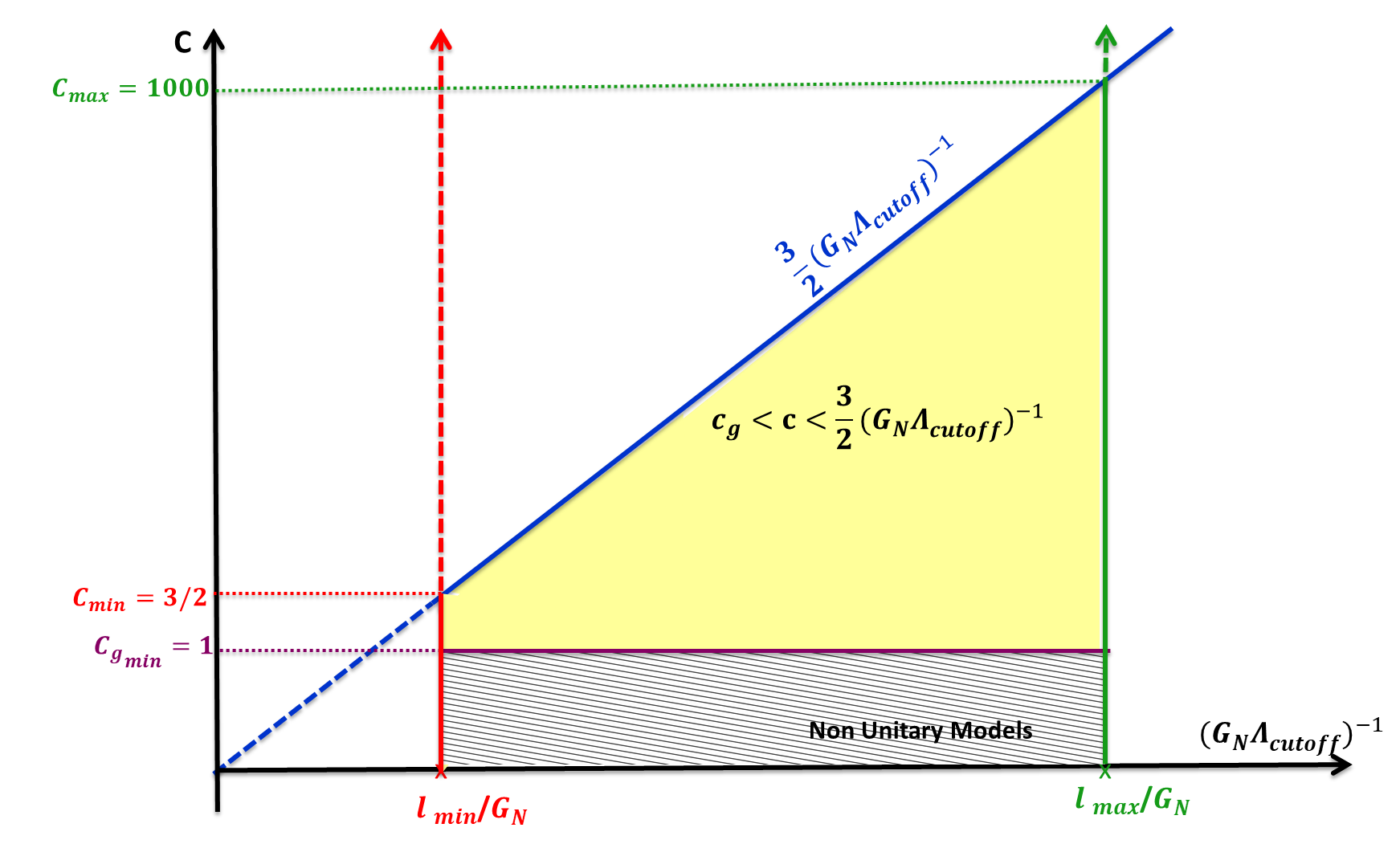}
\end{center}
\par
\vspace{-0.5cm}
\caption{Narain Landscape region (in yellow) represents the viable space of
consistent theories with central charge $c$ constrained by the cutoff as $%
\frac{3}{2}\left( G_{N}\Lambda _{cut-off}\right) ^{-1}$ (blue line). The red
and green lines indicate the minimal and maximal values of c determined by
the allowed range of the AdS radius $G_{N}\lesssim l_{AdS_{3}}\lesssim
600G_{N}$. The purple line denotes the central charge $c_{g}$ of the current
algebra and bounds the gray shaded area excluded by unitarity. For each c,
many values of $c_{g}=\frac{p+q-2}{2}$ are possible with a minimum value of $%
c_{g_{\min }}=1$ corresponding to a Narain CFT with $p=q=2$. }
\label{finit}
\end{figure}

The finiteness of the Landscape is ensured by imposing bounds on the cutoff
scale $\Lambda _{cut-off}$ which can be delineated as follows:

\begin{equation}
l_{\min }\leq \Lambda _{cut-off}^{-1}\sim l_{AdS_{3}}\leq l_{\max }
\label{B1}
\end{equation}

\textbf{Lower Bound:\newline
}For small values of the AdS$_{3}$ radius $l_{AdS_{3}}\ll l_{\min },$ the
spacetime curvature increases as the Ricci scalar scales as $R\sim
-6/l_{AdS_{3}}^{2}$. In this instance, gravitational effects become too
strong which leads to a breakdown of the EFT. Therefore, to preserve the
validity of the semi-classical description, it is necessary to only consider
finite diameters by imposing a lower bound on the AdS radius to ensure a
consistent definition of the EFT and its cutoff \cite{OV}.

\textbf{Upper Bound:\newline
}In contrast, the upper bound concerns a large AdS$_{3}$ radius. It is
necessary to prevent the radius from running to infinity to preserve the AdS
description; otherwise, the space will be quasi-flat. Since the holographic
AdS/CFT correspondence relies on the gravitational bulk to be curved, the
duality will become poorly defined as the radius grows large. Additionally,
an infinite radius could lead to the emergence of an infinite tower of
massless states which violates the finiteness conjecture requiring that the
number of massless modes must be finite \cite{Finit1}-\cite{Finit4}.

\textbf{Implied Bounds on the gravitational CS level }$k^{G}$\textbf{:}%
\newline

The geometrical constraint (\ref{B1}) can be reformulated in terms of the
gravitational CS level $k^{G}$ as follows:%
\begin{equation}
1/4\lesssim k^{G}\lesssim 166\qquad \Rightarrow \qquad G_{N}\lesssim
l_{AdS_{3}}\lesssim 666G_{N}  \label{B2}
\end{equation}%
Using the relation $k^{G}=l_{AdS_{3}}/(4G_{N}),$ and requiring a lower bound
on the AdS radius $l_{Pl}\lesssim l_{\min }$ in order to remain large in
comparison with the Planck length \cite{Pl}, one can translate the left hand
side of (\ref{B1}) into a condition on the lowest possible CS level $k^{G}.$
In Planck units $l_{Pl}\sim G_{N}$ \cite{Pl2}, one gets $k^{G}\gtrsim 1/4.$%
\newline
The derivation of the upper bound is based on leveraging results from the
conformal bootstrap program. In \cite{Bos1}, authors demonstrate how
systematic numeric bootstrap techniques can be applied to flavored partition
functions. In our context, we refer to this work primarily for its numerical
results indicating that holographic CFTs with large central charge $%
c=6k^{G}=3l_{AdS_{3}}/(2G_{N})$ and a sparse low-lying spectrum typically
yield central charges on the order of $c\sim 10^{3}.$ In \cite{Bos2},
similar bootstrap methods were applied with an emphasis on the presence of
gaps in the operator spectrum, sometimes assuming specific spin truncations.
These studies suggested an upper bound on the central charge consistent with
$c\sim 10^{2}.$ For $c\sim 10^{3},$ we estimate an upper bound on the CS
level given by $k^{G}\lesssim 166$ corresponding roughly to $l_{\max }\sim
666G_{N}.$

\section{Conclusion and discussion}

In this work, we have strengthened the connection between the Swampland
program and holography by focusing on AdS$_{3}$ gravity dual to an ensemble
of standard and generalised Narain CFTs. Prior studies have established
that, in order to define a consistent ensemble averaging over the Narain
moduli space, one must introduce fluctuations around the averaging to
prevent global symmetries from emerging \cite{M3}. These fluctuations were
shown to correspond, in the bulk, to a sublattice of super extremal
particles emitted by the discharging BTZ black holes consistent with the
weak gravity conjecture \cite{Fluc}. Building on this, we investigate how
additional Swampland criteria can further constrain the boundary ensemble to
ensure a consistent gravitational bulk.

For this purpose, we focused on the AdS distance conjecture and the
finiteness of the Landscape criteria seeing that in AdS settings, these
constraints are intertwined \cite{rev}. On one hand, the AdS distance
conjecture expects the emergence of a tower of light states as the AdS
radius increases. On the other hand, to prevent an unbounded number of such
theories, the finiteness conjecture requires a cutoff so the number of AdS$%
_{3}$/CFT$_{2}$ pairs is finite. We sharpened this latter requirement by
considering unitarity constraints which implied a condition on the rank of
gauge groups that can couple consistently to AdS$_{3}$ gravity (\ref{F}).
This, in turn, imposed an upper bound on the central charge of the CFT
current algebras implying a finite number of theories.

To compute this bound, we analyzed the emergent anomalies in the
gravitational bulk with respect to gravitational and gauge invariance (\ref%
{res}). We showed that the gravitational sector, composed of pure AdS$_{3}$
gravity, is anomaly free provided that one impose the Grumiller-Riegler
boundary conditions (\ref{bndy}) with equal left and right central charges.
The gauge sector, however, exhibits a more intricate behaviour. For the
standard Narain case, we only have gauge anomalies leading to two copies of
the $u(1)_{K}^{p-1}$ current algebra, each with central charge $c_{g}=p-1$.
In the generalised Narain case, in addition to the gauge anomalies enhancing
the symmetry by two copies of the $u(1)_{K}^{p-1}\oplus u(1)_{K}^{q-1}$
current algebra with distinct central charges $c_{g}=p-1$ and $\tilde{c}%
_{g}=q-1$, the sector also suffers from a gauge-sourced gravitational
anomaly stemming from an imbalance between the number of added U(1) gauge
fields to the left and right sectors.

Proceeding towards (\ref{F}), we provided mechanisms to cancel the anomalies
and ensure the consistency of the theory. For gauge anomalies, we exploited
the CS/WZW correspondence by introducing boundary strings as 2D CFTs with
gauge currents of opposite signs, effectively canceling the bulk
contributions. As for the gravitational anomalies, we added fermionic
degrees of freedom to compensate the mismatch between left and right gauge
central charges.

To illustrate the finiteness bound (\ref{F}), we gave an explicit
construction of the finite Narain Landscape by formulating realisations of
both the lower and upper bounds. We showed that by using results from
AdS/CFT, Swampland and conformal bootstrap program, one can compute concrete
limits on the values of the CS level $k^{G}$ and the corresponding AdS$_{3}$
radius (\ref{B2}). This allowed us in consequence to delineate the range of
the central charge and thus mapping the area of the Narain Landscape.

Altogether, these results provide new support for the validity of the
Finiteness conjecture and deepen our understanding of how the Swampland
criteria manifest in holographic settings. Our analysis also shows how
gravitational consistency can impose rigorous constraints on the boundary
ensemble, motivating further investigations into how other Swampland
criteria might help structuring the dual Landscape. Future work in this
direction will reveal deeper connections between Swampland and holography.

\section*{Acknowledgements}
 The work of R. Sammani is funded by the National Center for Scientific and Technical
 Research (CNRST) under the PhD-ASsociate Scholarship — PASS.
\section*{Conﬂict of Interest}
The authors declare no conﬂict of interest.
\section*{Data Availability Statement}
Data sharing not applicable to this article as no datasets were generated or analysed during the current study.

\end{document}